\documentclass{sig-alternate}
\usepackage{float}
\usepackage{color}
\usepackage{verbatim}
\usepackage{subcaption}
\usepackage{dblfloatfix}

\usepackage{etoolbox}% http://ctan.org/pkg/etoolbox
\makeatletter
\patchcmd{\maketitle}{\@copyrightspace}{\@float{copyrightbox}[b]\begin{center}\setlength{\unitlength}{1pc}\begin{picture}(20,2)\put(0,-0.95){\crnotice{\@toappear}}\end{picture}\end{center}\end@float}{}{}
\makeatother

\permission{Copyright is held by the author/owner(s).}
\conferenceinfo{WWW'16 Companion,}{April 11--15, 2016, Montr\'eal, Qu\'ebec, Canada.} 
\copyrightetc{ACM \the\acmcopyr}
\crdata{978-1-4503-4144-8/16/04. \\
http://dx.doi.org/10.1145/2872518.2890552}

\begin{document}

\title{Human Atlas: A Tool for Mapping Social Networks}
%
% You need the command \numberofauthors to handle the 'placement
% and alignment' of the authors beneath the title.
%
% For aesthetic reasons, we recommend 'three authors at a time'
% i.e. three 'name/affiliation blocks' be placed beneath the title.
%
% NOTE: You are NOT restricted in how many 'rows' of
% "name/affiliations" may appear. We just ask that you restrict
% the number of 'columns' to three.
%
% Because of the available 'opening page real-estate'
% we ask you to refrain from putting more than six authors
% (two rows with three columns) beneath the article title.
% More than six makes the first-page appear very cluttered indeed.
%
% Use the \alignauthor commands to handle the names
% and affiliations for an 'aesthetic maximum' of six authors.
% Add names, affiliations, addresses for
% the seventh etc. author(s) as the argument for the
% \additionalauthors command.
% These 'additional authors' will be output/set for you
% without further effort on your part as the last section in
% the body of your article BEFORE References or any Appendices.

\numberofauthors{4} %  in this sample file, there are a *total*
% of EIGHT authors. SIX appear on the 'first-page' (for formatting
% reasons) and the remaining two appear in the \additionalauthors section.
%
\author{
% You can go ahead and credit any number of authors here,
% e.g. one 'row of three' or two rows (consisting of one row of three
% and a second row of one, two or three).
%
% The command \alignauthor (no curly braces needed) should
% precede each author name, affiliation/snail-mail address and
% e-mail address. Additionally, tag each line of
% affiliation/address with \affaddr, and tag the
% e-mail address with \email.
%
% 1st. author
\alignauthor
Martin Saveski\\
\affaddr{MIT Media Lab}\\
\email{msaveski@mit.edu}
% 2nd. author
\alignauthor 
Eric Chu\\
\affaddr{MIT Media Lab}\\
\email{echu@mit.edu}
\and
% 3rd. author
\alignauthor 
Soroush Vosoughi\\
\affaddr{MIT Media Lab}\\
\email{soroush@mit.edu}
% 4th. author
\alignauthor 
Deb Roy\\
\affaddr{MIT Media Lab}\\
\email{dkroy@media.mit.edu}
}

\begin{comment}
\numberofauthors{1}
\author{
\alignauthor
Martin Saveski, Eric Chu, Soroush Vosoughi, Deb Roy\\
\affaddr{Laboratory for Social Machines, MIT Media Lab.}\\
\email{\{msaveski, echu, soroush\}@mit.edu, dkroy@media.mit.edu}
}
\end{comment}

\maketitle
\begin{abstract}
Most social network analyses focus on online social networks. While these networks encode important aspects of our lives they fail to capture many real-world connections. Most of these connections are, in fact, public and known to the members of the community. Mapping them is a task very suitable for crowdsourcing: it is easily broken down in many simple and independent subtasks. Due to the nature of social networks---presence of highly connected nodes and tightly knit groups---if we allow users to map their immediate connections and the connections between them, we will need few participants to map most connections within a community. To this end, we built the Human Atlas, a web-based tool for mapping social networks. To test it, we partially mapped the social network of the MIT Media Lab. We ran a user study and invited members of the community to use the tool. In 4.6 man-hours, 22 participants mapped 984 connections within the lab, demonstrating the potential of the~tool. 
\end{abstract}

% A category with the (minimum) three required fields
%\category{H.4}{Information Systems Applications}{Miscellaneous}
%A category including the fourth, optional field follows...
%\category{D.2.8}{Software Engineering}{Metrics}[complexity measures, performance measures]
%\terms{Delphi theory}
%\keywords{ACM proceedings, \LaTeX, text tagging}

\begin{figure*}[t]
    \centering
    \begin{subfigure}[t]{0.49\textwidth}
        \centering
        \includegraphics[width=\linewidth, trim={0cm 0cm 0cm 0cm}, clip]{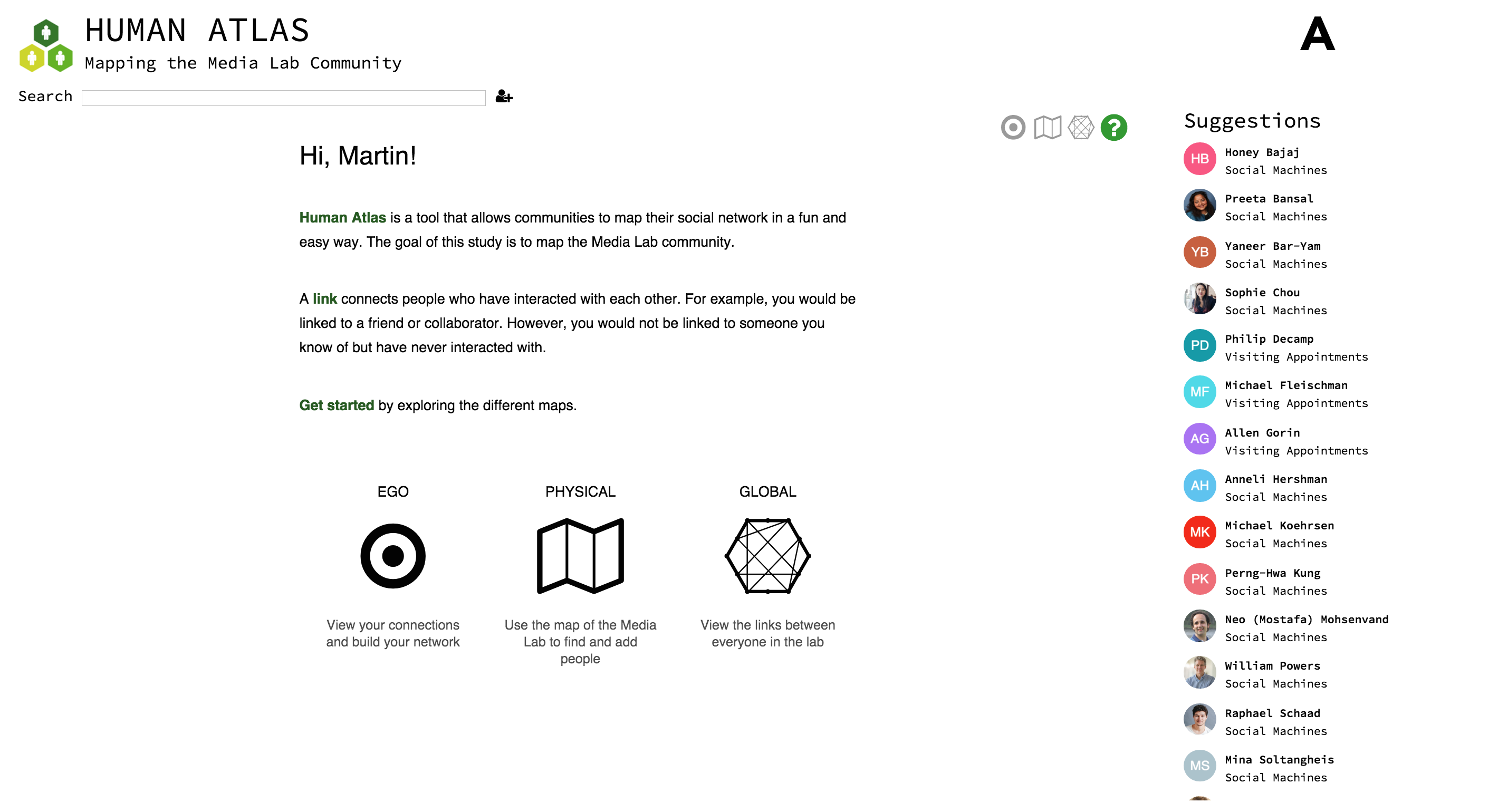}
    \end{subfigure}%
    \quad 
    \begin{subfigure}[t]{0.49\textwidth}
        \centering
        \includegraphics[width=\linewidth, trim={0cm 0cm 0cm 0cm}, clip]{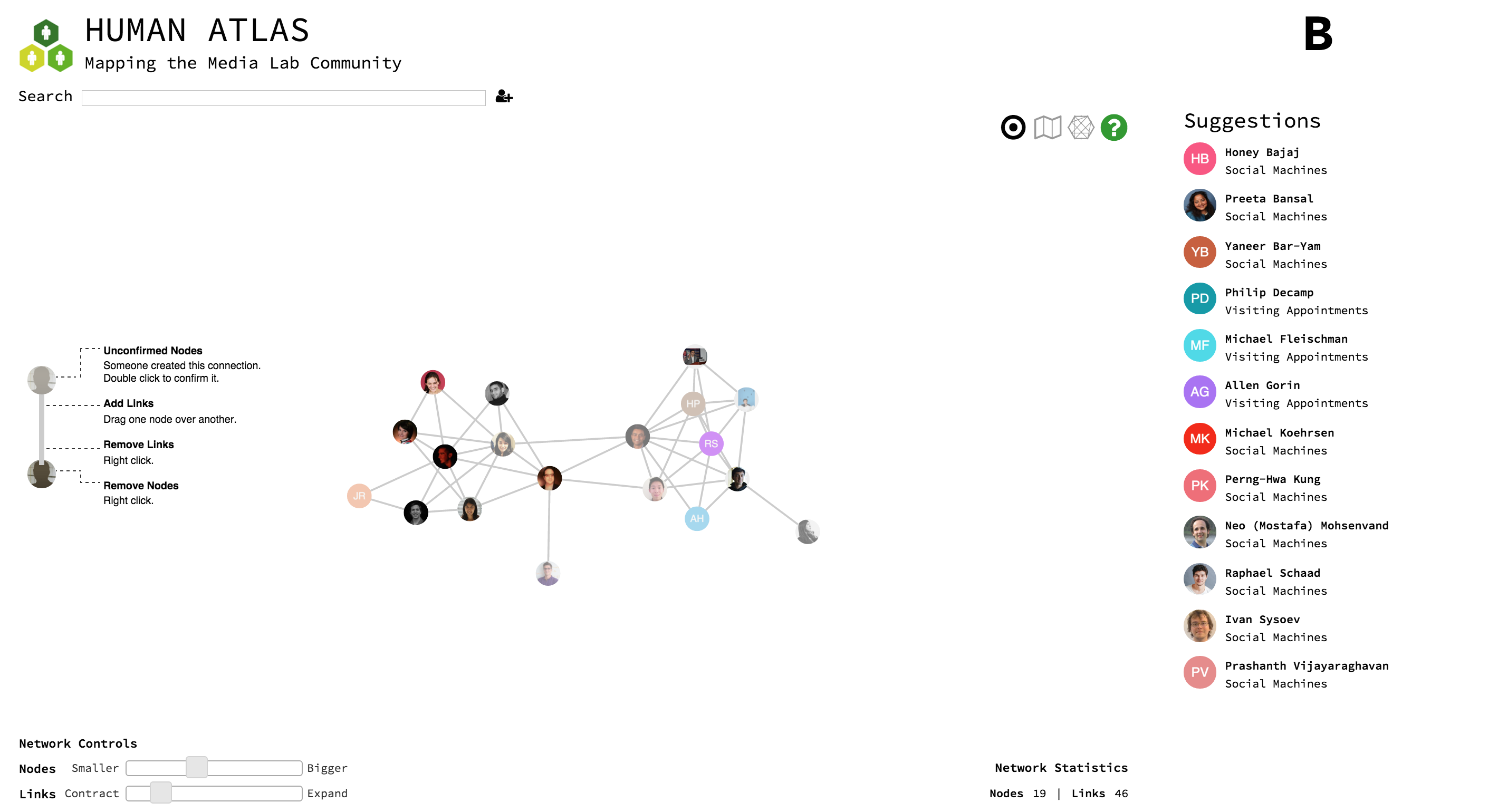}
    \end{subfigure}
    \vskip\baselineskip
    \begin{subfigure}[b]{0.49\textwidth}
        \centering
        \includegraphics[width=\linewidth, trim={0cm 0cm 0cm 0cm}, clip]{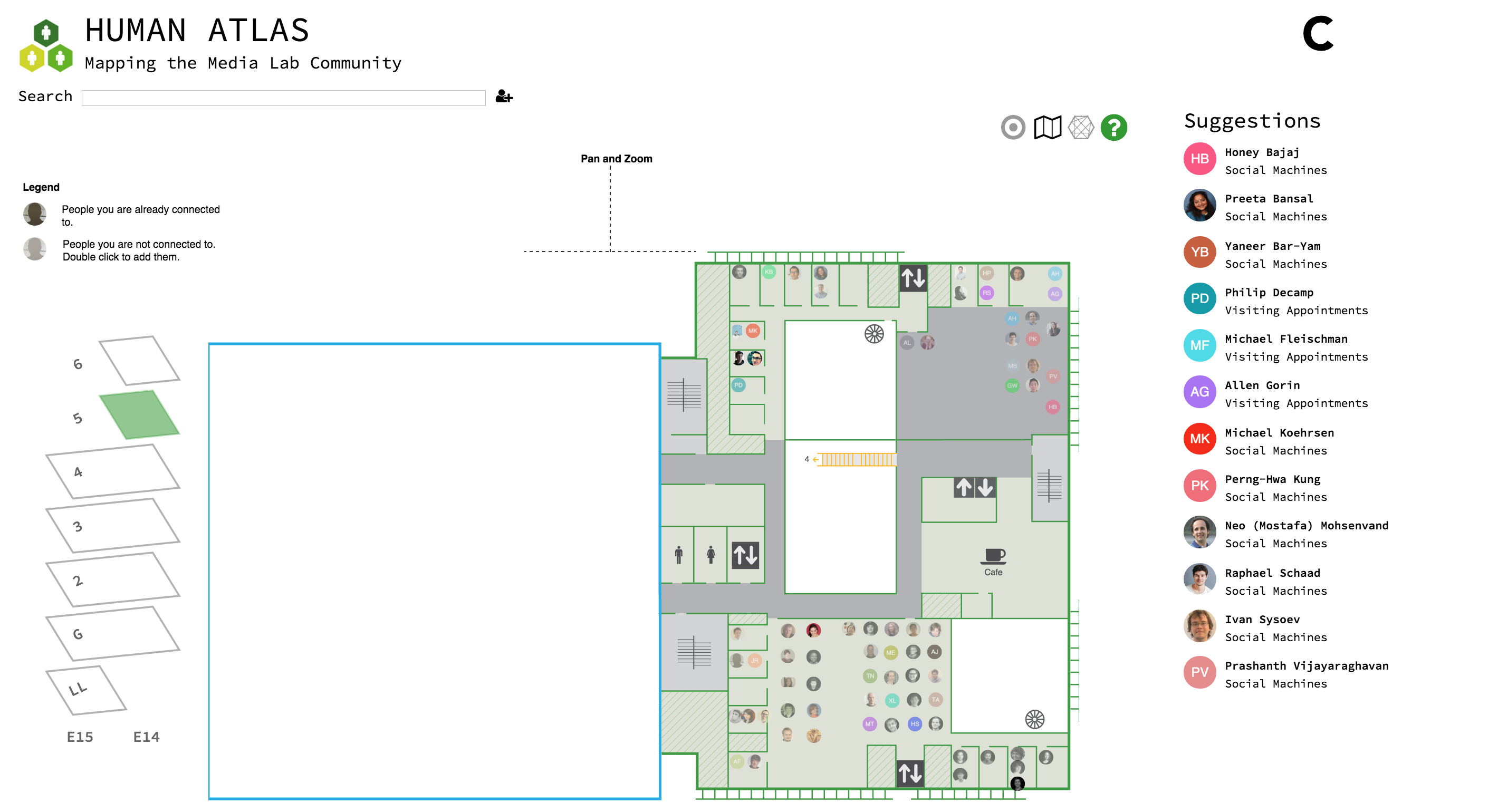}
    \end{subfigure}%
    \quad
    \begin{subfigure}[b]{0.49\textwidth}
        \centering
        \includegraphics[width=\linewidth, trim={0cm 0cm 0cm 0cm}, clip]{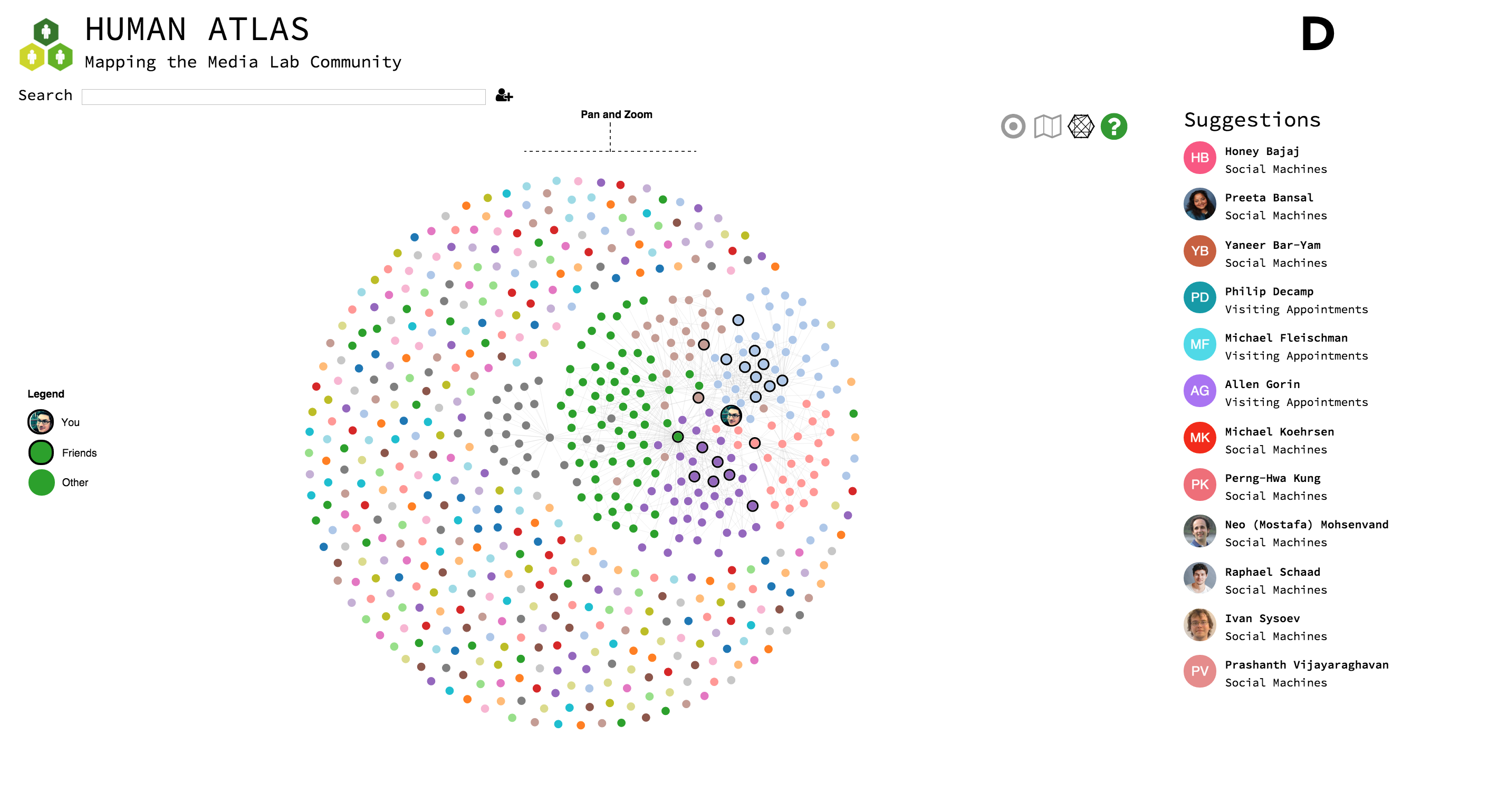}
    \end{subfigure}
    \caption{(A) Splash page: introduction to the Human Atlas. (B) Ego view: your immediate network. (C) Physical view: map of the Media Lab. (D) Global view: the entire community network.}
    \label{fig:HA-screens}
\end{figure*}

\section{Introduction}
Most social network analyses focus on online social networks, such as Facebook, Twitter, and LinkedIn. While these networks encode important aspects of our social lives, they fail to capture many important real-world connections. For instance, Facebook may capture connections to friends and family, while LinkedIn may capture professional ties. Moreover, most of these data are proprietary and the complete social networks are not available to the members of the community themselves.

However, most of these connections are public and known to many members of the community. They constitute, what we call, the \textit{publicly knowable graph}. Two people are connected in the publicly knowable graph if they interact with each other and also there are others who can confirm that they know each other. While different parts of the publicly knowable graph may be common knowledge among groups in the community, putting these fragmented pieces together may reveal important insights about the community. 

The task of mapping the publicly knowable graph is very well suited for crowdsourcing: ($i$) it is well-structured, and ($ii$) it has low complexity~\cite{franzoni2014crowd}. It is straightforward to divide and distribute it: we can ask every participant to map their immediate social network (i.e. ego network) and combine the individual networks to assemble the social network of the entire community. Moreover, the task of building an ego network is simple and does not require any specific skill. 
One example of a prior attempt to map publicly knowable connections through crowdsourcing for a specific domain is the Mathematics Genealogy Project, which aims to map the mentorship relationships between mathematicians. 

To maximize the number of connections captured during the mapping process, we have to enable participants not only to map their immediate connections, but also the connections between their connections. This allows us to capture more connections with much fewer participants. 

Social networks tend to exhibit properties of scale free networks~\cite{barabasi1999emergence}. The node degree distribution follows a power law, which indicates that highly connected individuals are very likely to occur and dominate the connectivity. Thus, if we have only a few highly connected individuals participating in the mapping process (mapping their immediate connection and the connections between them), we are very likely to capture a substantial fraction of all connections within the community.

Moreover, the distribution of clustering coefficients in social networks also follows a power law~\cite{watts1998collective}. This implies that people tend to create tightly knit groups characterized by a relatively high density of connections. Thus, by allowing participants to map the connections between their connections, we need very few individuals from a tightly knit group to map all connections within the group. Potentially, only one person is needed to report all members of the group and the connections within it. Finally, even if many individuals do not participate in the mapping process, due to the structure of the network, it is very likely that most of their connections will be mapped by others. 

This, in turn, requires a very specific interface design choice. Instead of just presenting users with a list of names of all the members of the community and allowing them to sort through and report who they know, we have to allow them to build a network and report the connections between their connections. 

With these ideas in mind, we set out to build the Human Atlas, a web application for mapping social networks. To test it, we decided to map the publicly knowable graph of our own community, the MIT Media Lab. We ran a user study and invited members of the community to use the tool and provide us feedback. In 4.6 man-hours, 22 participants mapped 984 connections within the lab. This demonstrates the power of the tool and its potential for immediate application.

\section{Human Atlas User Experience}
Next, we describe the Human Atlas user experience\footnote{Demonstration video is available at the following URL: https://youtu.be/OQOUHkJdA-U}. Screenshots of the main views are shown in Figure~\ref{fig:HA-screens}. \\ 
\vspace{-3.3mm}

\noindent
\textbf{Splash Page.} The splash page introduces the user to the concept of the publicly knowable graph and the purpose of the tool. It also defines a link as a connection between people who have interacted with each other. It then invites the user to start mapping the community by giving a short description of each of the different views. From the splash, the user can see the search bar and list of suggestions, which are the primary ways of adding people to one's network. These elements are constant across all four views both to establish visual consistency and to encourage the creation of new connections. Finally, three of the four icons to the left of the suggestions serve as a navigation menu between views, while the fourth acts as a view-dependent help button. \\
\vspace{-2.3mm}

\noindent
\textbf{Ego View.} The ego view is the primary interface for visualizing one's immediate network. The user's connections are displayed as nodes in a force-directed layout, in which the size of the nodes and length of the links can be changed to improve readability. Any links between connections will be present in the layout. Connections created by someone else are transparent until confirmation through double clicking. Simple statistics detailing the number of nodes and links in one's network are also provided. Finally, toggling the help highlights the fact that users can add links between two connections, a unique aspect of our tool that allows fewer people to fully map the community. \\
\vspace{-2.3mm}

\noindent
\textbf{Physical View.} As relationships often exist in physical spaces, the physical view provides a navigable map of the Media Lab. The panels on the left allow users to traverse between floors, with the ability to zoom and pan in each floor. Lab members are placed in their respective office locations, and a new connection can be added by double clicking on his or her avatar. This view can highlight the important role of physical proximity in which relationships we form. It can also act as a directory for finding where friends and collaborators sit. \\
\vspace{-2.3mm}
%When a connection is added through recommendation or search, the new node either floats to his or her location in the current floor, or it floats to the appropriate panel in the left floor navigation. \\

\noindent
\textbf{Global View.} Finally, the global view presents the current map of the entire network. The user's node is highlighted, and the user's first degree connections have bolded borders. Sub-communities within the global network are color-coded. When a connection is added through recommendation or search, a bold edge between the user and the new node is temporarily created. This view sheds insight into both the community structure and the user's place within it. 

%-----------------------------------------------------------
\section{Implementation Details}
\noindent
\textbf{Algorithms.}
We use \textit{triadic closure} recommendations~\cite{carullo2015triadic} to generate a list of people who the user may interact with. The main intuition behind triadic closure is that if two people have many connections in common then they are very likely to be connected. The recommendation engine sorts people by the number of shared connections with the people who are already in the user's network---the more mutual connections, the higher a person is ranked in the list of suggestions. To bootstrap the recommendations, we initially suggest people from the user's research group.

To minimize the visual clutter when displaying the networks in the ego and global view, we color densely connected nodes in the same color. We apply the Louvain \textit{community detection} algorithm~\cite{Blondel2008} to find groups of nodes that have many connections between each other, but are less connected to the other nodes in the graph. The main advantages of this algorithm are both the automatic detection of the optimal number of communities (no need to set that number a priori) and its high clustering accuracy~\cite{fortunato10}. \\
\vspace{-2.3mm}

\noindent
\textbf{Technology.}
The backend was built in Python and deployed as a web application using the Flask framework. For the front end visualizations, we used the Javascript library D3. Of note is the force-directed graph layout used for the Ego and Global views, which represents a network through a physical simulation of charged particles and springs connecting them. This is parameterized by the magnitude of the repulsive charge and the magnitude of a gravity force that keeps nodes centered in the visible area. 

%-----------------------------------------------------------
\section{Design Considerations}
\noindent
\textbf{Definition of a Link.} 
We aimed to provide a definition that would capture meaningful relationships, but not one so restrictive that we would fail to capture the ``true'' social network. We settled on links connecting people who have interacted with each other. For example, one would be linked to a friend or collaborator. However, one would not be linked to someone he or she knows of but has never interacted with.  \\
\vspace{-2.3mm}

\noindent
\textbf{Heterogeneous Links.}
Even armed with our definition, we considered the possibility of heterogeneous links. For example, links could have different strengths on a scale of one to five, or they could explicitly specify the nature of the relationship. The latter could help users find potential research connections, while this general scheme would be useful for analyzing the graph and comparing to different social networks.  Ultimately, however, we wanted the tool to be as low friction as possible. Having to specify the link type every time could strongly inhibit engagement. We also believed that defining the nature of each link could give rise to valid privacy concerns. However, we have implemented link types in the backend in anticipation that this tool can be adapted for different communities. \\
\vspace{-2.3mm}

\noindent
\textbf{Confirmation/Privacy of Links.}
To address privacy concerns about what is publicly knowable, links can be confirmed on either end. When a link has been added by someone else, the recipient of that link has the option to confirm it on his or her side. Unconfirmed links are shown in the Ego view as transparent nodes. If a link has not been confirmed by both sides, it will not show up in someone else's ego network. We call this \textit{privacy by default}. Currently, however, these edges do show up in the global view. While our user study survey results indicated that privacy concerns were not high, our design will allow us to toggle these public edges as needed. \\
\vspace{-2.3mm}

\noindent
\textbf{Ego view.}
Force layouts quickly become visually cluttered when the network is too large. To help counteract this, we first chose not to show the user in his or her own graph. Given that this is a view of the ego network, edges to the nodes in the graph are implicit. We also provided sliders to adjust the repulsive charge and size of the nodes. When the nodes become very small, the pictures within disappear and only small colored dots remain. This allows users to gain an understanding of the overall clustering and topology of their network. 
%In the future, we plan to use a regression model to automatically set the charge and size as a function of the size and structure of the network, thereby lifting the burden of fine-tuning the visualization from the users. 

%-----------------------------------------------------------
\section{User Study}
\noindent
\textbf{Setup.} To understand how users use and perceive the Human Atlas we ran a user study. We invited 29 members of the Media Lab community to try the tool and give us their feedback. The participants were mostly graduate students, with a few faculty and research staff included, and belonged to 9 (out of 25) different research groups. We made sure that there was a mixture of participants from different academic backgrounds, as well as people who have been a part of the lab for different amounts of time. 

We informed the participants about the study by sending a short invitation email that gave a one-sentence description of the tool and provided links to the web interface and survey. To test whether the tool is intuitive and easy to use, we purposefully gave no instructions. The survey included nine questions (mostly multiple-choice) asking participants' opinions about the different views, quality of the recommendations, privacy concerns, and whether they found anything confusing. To gain better insight in how the participants used the tool, we logged the most important user actions: switching views, searching for people, and adding nodes and links. The participants were given five days to try the tool and complete the survey. The participation rate was 75\% for using the tool (22 out of 29), and 52\% for completing the survey (15 out of 29). \\
\vspace{-2.3mm}

\begin{figure}[t]
\centering
\includegraphics[width=\linewidth, trim={25mm 110mm 123mm 85mm}, clip]{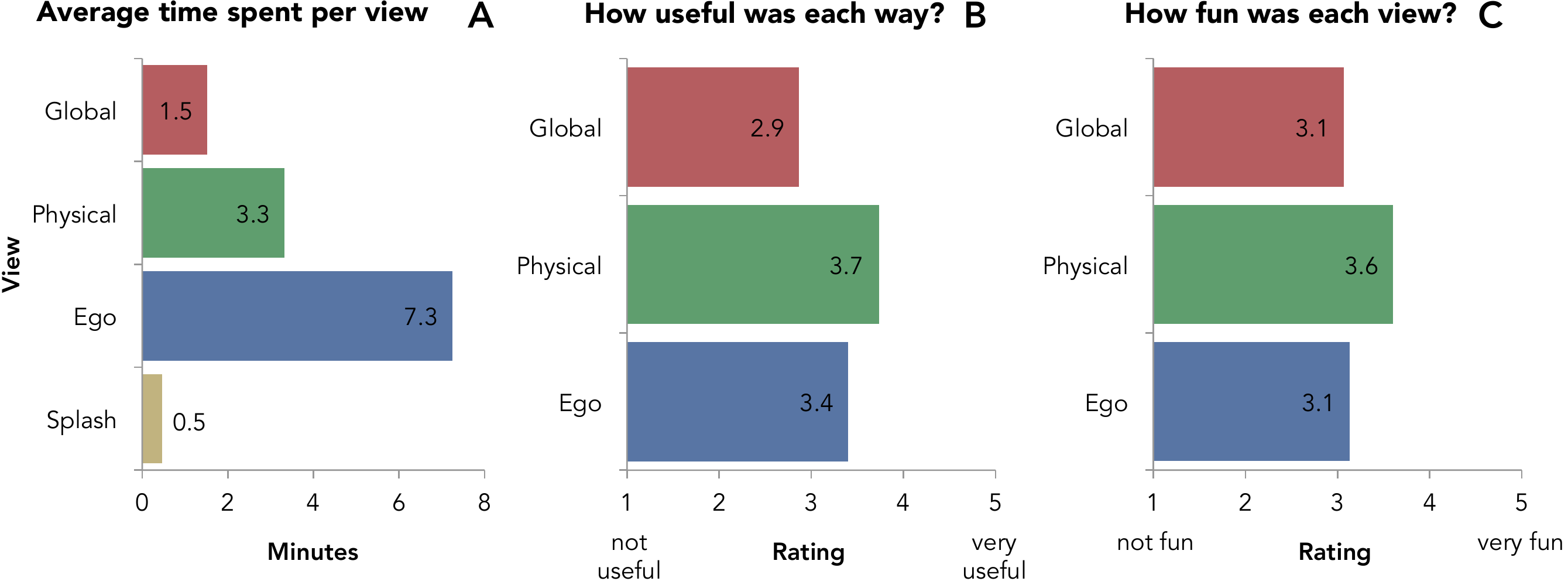}
\caption{(A) Average time spent per view. (B-C) Survey results, average rating of how useful (B) and fun (C) each view was.}
\label{fig:views}
\end{figure}

\noindent
\textbf{Views.} 
The average session was 6.7 minutes, and most participants had only one session. Users spent the most time on the ego view at 7.3 minutes/user, followed by the physical view at 3.3 minutes/user, and finally the global view at 1.5 minutes/user (Figure~\ref{fig:views}A). This is most likely due to the design of the tool, as the ego view provides the most features and is a natural first stop in the user flow. However, when asked to rate the different views on a scale from 1 (not useful/fun) to 5 (very useful/fun), the majority found the \textit{physical view to be the most useful and fun view}. As shown in Figure~\ref{fig:views}B, 46\% found the physical view to be the most useful view, with an average rating of 3.5. This was followed by the ego view with 34\% and an average rating of 3.4, and then the global view with 20\% and an average rating of 2.9. Similarly, 42\% of people found the physical view to be most fun, with an average rating of 3.6 (Figure~\ref{fig:views}C). \\
\vspace{-2.3mm}

\begin{figure}[t!]
\centering
\includegraphics[width=\linewidth, trim={18mm 52mm 47mm 78mm}, clip]{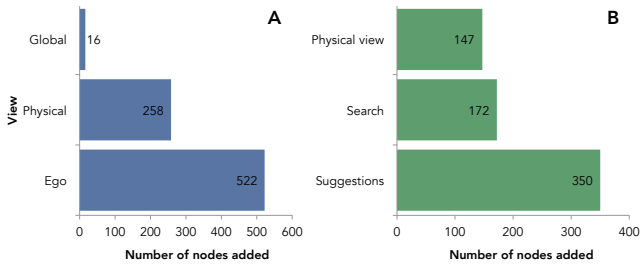}
\caption{(A) Number of nodes added per view. (B) Ways of finding the immediate connections.}
\label{fig:nodes}
\end{figure}

\noindent
\textbf{Adding immediate links.}
To add nodes to their network, participants mostly used the ego view, with 65\% of all immediate connections added from this view (Figure~\ref{fig:nodes}A). This was followed by the physical view, from which 32\% of the immediate connections were added. The global view was almost not used for this purpose. Again, this is most likely due to the deign of the tool. Very few links were deleted, with most deletions occurring right after they were created. This indicates that link deletion is being used as an undo feature.

To find others and add them to their network, participants can use the search box, the list of suggestions, and the physical map of the lab. As shown in Figure~\ref{fig:nodes}B, most immediate connections were added from the list of suggestions (52\%), followed by search (26\%), and the physical view (22\%). This aligns with the feedback provided in the survey. When asked to rate the quality of the suggestions on a scale from 1 (not relevant) to 5 (very relevant), 80\% of the participants gave a rating of 3 or higher, with average rating of 3.4. \\
\vspace{-2.3mm}

\noindent
\textbf{Third-party links.} One of the key advantages of the Human Atlas is that users can add links between their immediate connections. However, before running the user study, we were uncertain whether users would be interested and willing to add these links. To our surprise, 39\% of all links captured were links between users' immediate connections. Some found it fun, with one stating ``It seemed nice to see how my actual ego network would look like". \\
\vspace{-2.3mm}

\noindent
\textbf{Confirming links.}
When a user adds a new connection, only creator's end of the link is confirmed. The user on the other end may confirm the link or delete it. If a link is introduced by a third party (a common connection), then both ends of the link are unconfirmed. We found that 10\% of the links were confirmed, out of all the links that could have been confirmed given the users who participated in the study. The reason for this may be the fact that most users used the tool only once, and many of their links were not created when they used it. One way to overcome this is to send users a notification when a link to them is created. Many users who did not participate in the study were also added.\\
\vspace{-2.3mm}

\noindent
\textbf{Privacy concerns.}
We asked the participants to rate their privacy concerns about the Human Atlas on a scale from 1 (not concerned at all) to 5 (very concerned). Nearly half (47\%) were not concerned at all, and the average rating was 1.9. Some considered the information captured by the tool public, and elaborated ``This is public information'', or ``I felt like this info is available anyways.'' Others were more concerned, stating ``[I] found myself hesitant to map my own social connections in a public way. Didn't feel like my kind of thing.'' We also asked participants to rate on a scale from 1 to 5 if they thought that others in the Media Lab community would have privacy concerns. More than half of them (54\%) thought that others would be more concerned than them, and the average rating was 2.7. \\
\vspace{-2.3mm}

\noindent
\textbf{General feedback.}
Some participants found it confusing that their node is not in their ego network. Others were not clear about the semantics of the links and were unsure whether to add some links or not. A few participants mentioned that it would be useful to be able to refresh the list of suggestions or mark some suggestions as irrelevant. 

\section{Demonstration plan}
The audience will be given the opportunity to fully explore the Human Atlas. The goal of the demonstration will be to showcase the tool by allowing the attendees to map the publicly knowable graph of WWW'16. 

Before the conference, we will pre-populate the database with names and co-authorship relationships by scraping the ``accepted papers'' section of the WWW'16 conference website. During the conference, attendees whose name is not in the database will be given a chance to manually enter their name into the system.

All users will also be given the opportunity---but will not be required---to upload their pictures to the tool so that it can be shown in the graph. The complete tool, including its various views and algorithms will be available for the users to explore. The only notable exception is the physical view, which we cannot demonstrate since it is unsuitable for a conference setting where people do not have permanent locations. 

As more attendees use the tool, the publicly knowable graph of the conference attendees should become more complete and reflect the underlying social graph of the conference.

\section{Future Work}
We plan to add features to allow users to: ($i$) import connections from social media (e.g. Twitter); ($ii$) mark link suggestions as irrelevant and refine recommendations; ($iii$) add research interest tags and find potential collaborators. Finally, once we capture the complete network, we are interested in comparing it with online social networks, such as those captured by Twitter or Facebook.

% You must have a proper ".bib" file
%  and remember to run:
% latex bibtex latex latex
% to resolve all references
%
% ACM needs 'a single self-contained file'!
%
%\balancecolumns

\begin{thebibliography}{1}

\bibitem{barabasi1999emergence}
A.-L. Barab{\'a}si and R.~Albert.
\newblock Emergence of scaling in random networks.
\newblock {\em Science}, 1999.

\bibitem{Blondel2008}
V.~D. Blondel, J.-L. Guillaume, R.~Lambiotte, and E.~Lefebvre.
\newblock Fast unfolding of communities in large networks.
\newblock {\em Journal of Statistical Mechanics: Theory and Experiment}, 2008.

\bibitem{carullo2015triadic}
G.~Carullo, A.~Castiglione, A.~De~Santis, and F.~Palmieri.
\newblock A triadic closure and homophily-based recommendation system for
  online social networks.
\newblock {\em World Wide Web Conference}, 2015.

\bibitem{fortunato10}
S.~Fortunato.
\newblock Community detection in graphs.
\newblock {\em Physics Reports}, 2010.

\bibitem{franzoni2014crowd}
C.~Franzoni and H.~Sauermann.
\newblock Crowd science: The organization of scientific research in open
  collaborative projects.
\newblock {\em Research Policy}, 2014.

\bibitem{watts1998collective}
D.~J. Watts and S.~H. Strogatz.
\newblock Collective dynamics of `small-world' networks.
\newblock {\em Nature}, 1998.

\end{thebibliography}
\end{document}